\documentclass[aps,prl,reprint,showpacs,superscriptaddress,10pt]{revtex4-1}
\usepackage{amsmath,amssymb,amsfonts,graphicx,bm,longtable,dcolumn,color}
\usepackage{mathrsfs}
\usepackage[utf8]{inputenc}
\usepackage{textcomp}
\usepackage[T1]{fontenc}
\usepackage{lmodern}

\begin{document}

\title{Thermomechanical bistability of phase-transition oscillators\\driven by near-field heat exchange}

\author{Marta Reina}
\email{marta.reina@institutoptique.fr}
\affiliation{Laboratoire Charles Fabry, UMR 8501, Institut d'Optique, CNRS, Universit\'{e} Paris-Saclay, 2 Avenue Augustin Fresnel, 91127 Palaiseau Cedex, France}

\author{Riccardo Messina}
\affiliation{Laboratoire Charles Fabry, UMR 8501, Institut d'Optique, CNRS, Universit\'{e} Paris-Saclay, 2 Avenue Augustin Fresnel, 91127 Palaiseau Cedex, France}

\author{Svend-Age Biehs}
\affiliation{Institut f\"{u}r Physik, Carl von Ossietzky Universit\"{a}t, D-26111 Oldenburg, Germany}

\author{Philippe Ben-Abdallah}
\affiliation{Laboratoire Charles Fabry, UMR 8501, Institut d'Optique, CNRS, Universit\'{e} Paris-Saclay, 2 Avenue Augustin Fresnel, 91127 Palaiseau Cedex, France}

\begin{abstract}
Systems with multistable equilibrium states are of tremendous importance in information science to conceive logic gates. Here we predict that simple phase-transition oscillators driven by near-field heat exchanges have a bistable thermomechanical behavior around their critical temperature, opening so the way to a possible boolean treatment of information from heat flux at microscale.
\end{abstract}

\maketitle

Processing information with heat rather than with electrons to perform logical operations is a challenging problem in modern physics. Several directions have been explored during the last decade to this end. In 2004 Li et al.~\cite{Li1} have demonstrated the possibility to locally break the symmetry for the phonon propagation inside non-linear atomic lattices opening so the door to the control of heat flow in solid-state elements and consequently to a possible logical treatment of information carried by thermal phonons~\cite{Chang,Li2,Li3}. More recently, to overcome the inherent problems associated with the relatively small propagation speed of acoustic phonons, solid photonics systems have been proposed to achieve similar operations~\cite{pba1,pba2,pba3} in contactless many-body systems~\cite{pba4,Riccardo1,Riccardo2,Ivan,Zhu2016,EkerothEtAl2017,MuellerEtAl2017,ZhuEtAl2018}. Finally, heat transport mediated by spin waves has also been considered~\cite{Poletti} to rectify heat flux in many-body systems, paving thus the way to highly-performing quantum devices for thermal computing.

\begin{figure}
	\includegraphics[width=0.47\textwidth]{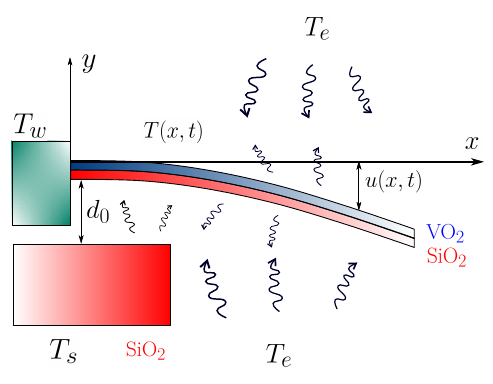}
	\caption{Sketch of thermomechanical oscillator. A bilayered cantilever is supported at its left end by a wall at temperature $T_w$, being kept free at the other end. It interacts through radiative heat transfer with an environment at temperatures $T_e$ and in near field with a substrate at temperature $T_s$. Its displacement is described by $u(x,t)$, while $T(x,t)$ represents its temperature profile.}
	\label{fig:sin}
\end{figure}

In this Letter we study a thermomechanical oscillator composed by a bilayer beam made of two different materials. Such a system has been experimentally studied in \cite{Kwon}, where the dynamic response to a periodic far-field heating was considered. We consider a beam made of a metal-insulator transition (MIT)~\cite{Mott} material and a dielectric, in interaction in near-field regime~\cite{van Zwol1,van Zwol2,Gotsmann,Reddy} with a substrate and in far field~\cite{VO21,VO22} with a thermal bath. When the equilibrium temperature along the beam is close to the critical temperature $T_c$ of the MIT material, we predict the existence of a bistable behavior. The system is depicted in Fig.~\ref{fig:sin}. The beam is recessed in a wall maintained at temperature $T_w$ by an external power source, whereas its right end is left free to oscillate. 

This cantilever exchanges heat radiatevely with an environment at temperature $T_e$ and in near field with a substrate at temperature $T_s$. Moreover, it is made of two different materials having thicknesses ($y$ axis in Fig.~\ref{fig:sin}) $h_1$ and $h_2$ ($h = h_1+h_2$). Its length ($x$ axis in Fig.~\ref{fig:sin}) and width ($z$ axis) are $L$ and $\delta$, respectively. The bottom layer of the cantilever is made of silicon dioxide (SiO$_2$)~\cite{Palik}, while the upper one of vanadium dioxide (VO$_2$) which undergoes a first-order transition at $T_c = 340\,$K~\cite{Barker}. The substrate is made of SiO$_2$ and has length $l$. The time evolution of the cantilever displacement $u(x,t)$ and temperature profile $T(x,t)$ is governed by the coupled system of nonlinear differential equations
\begin{equation}
\rho\, C h\, \partial_t T(x,t) = h \kappa\, \partial^2_x T(x,t) + \Phi\bigl(u(x,t), T(x,t)\bigr),\\
\label{Tuxt1}
\end{equation}
\begin{equation}
EI\,\partial^4_x u(x,t) = -\mu\,\partial^2_t u(x,t) - \gamma\, \partial_t u(x,t)- \partial^2_x M_T(T(x,t)),
\label{Tuxt2}
\end{equation}
corresponding to the energy-balance and Euler-Bernoulli~\cite{Roark} equations, respectively. Here $\rho$, $C$ and $\kappa$ are the beam mass density, specific heat capacity and thermal conductivity, while $\Phi \delta dx$ is the energy received (from the substrate and the far-field environment) per unit time by the infinitesimal element of the beam between $x$ and $x+dx$. $EI$ denotes the beam flexural rigidity, $\mu=\rho h\delta$ its linear mass density, $\gamma$ its damping and $M_T$ the thermal moment. Concerning the boundary conditions, we start by fixing the beam temperature on its recessed end [$T(0,t) = T_w$], while adiabatic conditions are applied on its right end [$\partial_{x}T(L,t) = 0$]. As for the initial thermal conditions, we assume a given temperature profile $T(x,0) = f(x)$. Concerning the displacement, we also impose a given profile at $t=0$, $[u(x,0) = g(x)]$, zero displacement and derivative at the fixed left end $[u(0,t) = \partial_x u(0,t) = 0]$ describing the wall support, zero velocity of any element of the beam $[\partial_t u(x,0) = 0]$, external bending moment at $x=L$ equal to the thermal moment $[\partial^2_x u(L,t) = M_T(T(L,t))]$~\cite{ZhangAIAA}, and zero third derivative at $x=L$ associate with the absence of a shear force $[\partial^3_x u(L,t) = 0]$~\cite{Kida}.

For a bimaterial cantilever, the mass dentity $\rho$, specific heat capacity $C$, thermal conductivity $\kappa$ and flexural rigidity $EI$ need to be replaced by effective quantities. More specifically, if the two parts of the beam have Young's moduli $E_1$ and $E_2$, densities $\rho_1$ and $\rho_2$ and thermal conductivities $\kappa_1$ and $\kappa_2$, $\rho$, $C$ and $\kappa$ are defined in terms of the following weighted averages
\begin{equation}
a = \dfrac{a_1 V_1 + a_2 V_2}{V_1 + V_2},\hspace{1cm}C = \dfrac{C_1 \rho_1 V_1 + C_2 \rho_2 V_2}{\rho_1 V_1 + \rho_2 V_2},
\label{w-a}\end{equation}
where $a\in\{\rho,\kappa\}$. The effective flexural rigidity $EI$ can be written as~\cite{Roark,Canetta}
\begin{equation}
EI = \dfrac{\delta h_1 h_2}{12} \dfrac{E_1 h_1 E_2 h_2}{E_1 h_1 + E_2 h_2} K,
\label{EI}\end{equation}
with
\begin{equation}
K = 6 + 4\biggl(\dfrac{h_1}{h_2} + \dfrac{h_2}{h_1} \biggr)+\biggl(\dfrac{E_1 h_1^2}{h_2^2} + \dfrac{E_2 h_2^2}{h_1^2} \biggr).
\label{K}\end{equation}
Finally, $M_T$ is the thermal moment that can be expressed as a function of the beam construction temperature $T_0$ (the one at which the beam is not curved), its temperature profile $T(x,t)$ and the thermal expansion coefficients $\alpha_1$ and $\alpha_2$ of the two materials, in the following way~\cite{ZhangAIAA}
\begin{equation}
M_T(T(x,t))=6 EI \dfrac{ (\alpha_2-\alpha_1)}{K}\biggl(\dfrac{1}{h_1}+\dfrac{1}{h_2}\biggr) (T(x,t)-T_0).
\label{MT}\end{equation}

The radiative heat flux is described within a fluctuational-electrodynamics approach, where the statistical properties of the charges fluctuating inside each body are accounted for by means of the fluctuation-dissipation theorem~\cite{Rytov}, correctly describing the heat exchange both in far and near field. In this framework, for small temperature differences the distance-dependent heat flux per unit surface between two planar substrates at distance $d$ and temperatures $T$ and $T+\Delta T$, respectively, can be written as $\Phi(d,T) = G(d,T)\Delta T$, where the conductance $G(d,T)$ is given by a sum over all the frequencies $\omega$, lateral wavevectors $\mathbf{k}$ and polarizations $p = \{\text{TE,TM}\}$ of the electromagnetic field as
\begin{equation}
G(d,T)=\int\limits_0^{\infty}\frac{\mathrm d\omega}{2\pi}\Theta'(\omega,T)\sum_p \int\limits_0^{\infty}\frac{\mathrm d^2\mathbf{k}}{(2\pi)^2} \mathcal T_p(\omega,k,d),
\end{equation}
where $\Theta'(\omega,T)$ is the $T$-derivative of the average thermal energy of a harmonic oscillator $\Theta(\omega,T)=\hbar\omega/(\exp(\hbar\omega/k_BT)-1)$. The energy transmission coefficient $\mathcal T_p(\omega,k,d)$ between two planar bodies 1 and 2 can be expressed in terms of the Fresnel coefficients $r_{ip}$ and $t_{ip}$ (reflection and transmission coefficients for body $i$ and polarization $p$) as~\cite{SurfSciRep}
\begin{equation}
\mathcal T_p(\omega,k,d)=\begin{cases}
\frac{(1-|r_{1p}|^2-|t_{1p}|^2)(1-|r_{2p}|^2-|t_{2p}|^2)}{|D_p|^2}, & ck < \omega,\\
\frac{4\,\text{Im}(r_{1p})\text{Im}(r_{2p})e^{-2|k_z|d}}{|D_p|^2}, & ck > \omega,
\end{cases}
\end{equation}
$k_z$ being the $z$ component of the wavevector and $D_p=1-r_{1p}r_{2p}e^{2ik_zd}$ a Fabry-P\'{e}rot denominator. In order to simplify the solution of the system of nonlinear differential equations \eqref{Tuxt1}-\eqref{Tuxt2}, we replace the conductance $G(d,T)$ by a polynomial expansion with respect to the separation distance 
\begin{equation}
G(d,T)=A_0 + A_1 d^{-1}+A_2 d^{-2}, 
\end{equation}
where $A_0 = 10.9\,\text{W}\,\text{m}^{-2}\,\text{K}^{-1}$, $A_1 = 0\,\text{W}\,\text{m}^{-1}\,\text{K}^{-1}$ and $A_2= 3.61 \times 10^{-12}\,\text{W}\,\text{K}^{-1}$  are the fitting parameters calculated from the exact expression of heat flux. 

\begin{figure}
	\hspace{-.15cm}\includegraphics[width=0.48\textwidth]{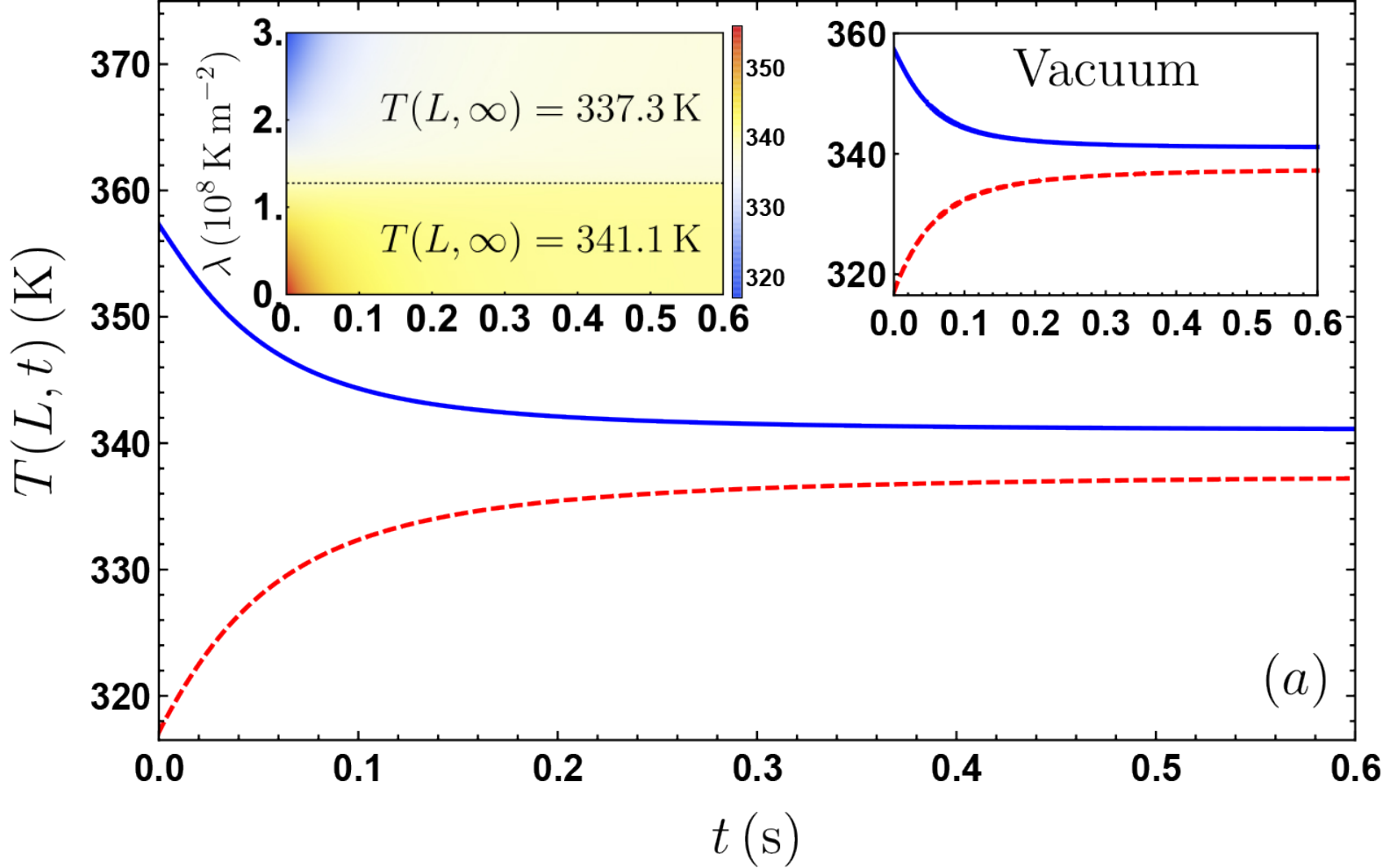}
	\includegraphics[width=0.47\textwidth]{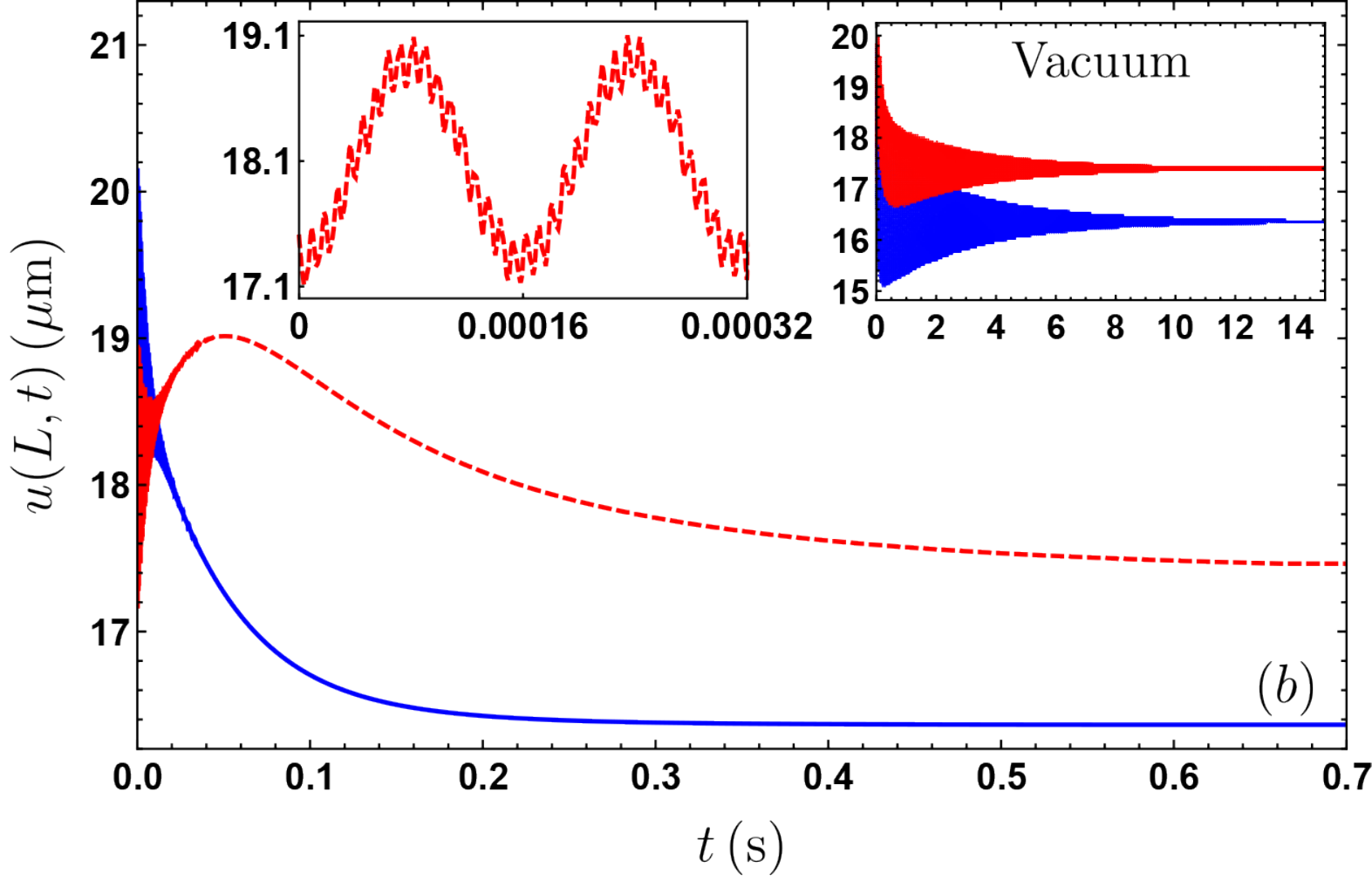}
	\caption{Time evolution of the (a) temperature and (b) displacement of the free end of the cantilever. The main part of the both plots refers to an evolution in air, the right insets in vacuum. The red-dashed curves are the lower-temperature solutions obtained for $\lambda = -0.1\times10^8$\,K\,m$^{-2}$, the blue-solid curves the upper-temperature solutions obtained for $\lambda = 3\times10^8$ K m$^{-2}$. The left inset of panel (a) describes the time evolution of $T(L,t)$ as a function of the parameter $\lambda$ associated to the initial temperature profile (see main text for details). The left inset of panel (b) gives the time evolution of $u(L,t)$ in air over a shorter timescale.}
	\label{fig:t-Tu}
\end{figure}

In order to demonstrate the existence of bistability we investigate the spatio-temporal evolution of displacement and temperature profiles by solving numerically the differential system \eqref{Tuxt1}-\eqref{Tuxt2}. The complexity of this system does not allow to derive analytically an existence criteria of bistability. However, it is clear that the equilibrium temperature of the beam must be close to the critical temperature of the MIT material. Therefore the temperatures of substrate, wall and thermal bath must not be all above or below $T_c$. To solve numerically the system of non-linear coupled differential equation we use a finite-difference method based on a implicit scheme~\cite{Dewey, Stab}. Concerning the geometric parameters, we choose values ($\delta = 1\,\mu$m, $L = 360\,\mu$m, $h_1=90$\,nm for the VO$_2$ layer, $h_2 = 910$\,nm for the SiO$_2$ layer, whereas the length of the substrate is $l=30\,\mu$m, placed at distance $d_0=200\,$nm from the $x$ axis) to ensure a relative balance between the magnitude of heat flux exchanged in far and near field. For SiO$_2$ we use $E_2=68\,\text{GPa}$, $C_2=730\,\text{J}\,\text{K}^{-1}\,\text{kg}^{-1}$, $\rho_2=2650\,\text{kg}\,\text{m}^{-3}$, $\alpha_2=8\times10^{-6}\,$K$^{-1}$, and the optical data given in Ref.~\cite{Palik}. For VO$_2$, we take $C_1=344\,\text{J}\,\text{K}^{-1}\,\text{kg}^{-1}$, $\rho_1=4570\,\text{kg}\,\text{m}^{-3}$ and $E_1=85\,\text{GPa}$ since it does not vary significantly with the phase~\cite{E1}. In order to describe the temperature dependence of the physical property $a_1$ of VO$_2$ [$a_1\in\{\alpha_1,\kappa_1,\epsilon_1\}$, $\epsilon_1$ being the emissivity] around $T_c$ we use the smoothing function 
\begin{equation}
S(T,T_c,\beta)=\dfrac{1}{1 - e^{- 2 \beta (T - T_c)}},
\label{d-m}\end{equation}
where $\beta$ is a parameter allowing to adjust the smoothness in the transition region between the dielectric and the metallic phases. In terms of this function, the physical quantity $a_1$ is given by
\begin{equation}
a_1(T)=a_{1d} +S(T,T_c,\beta)(a_{1m} - a_{1d}),
\label{transition}\end{equation}
where $a_{1d}$ and $a_{1m}$ are the properties in the dielectric and metallic phase, respectively. For film thicknesses of few dozens of nm the transition generally occurs over a temperature interval of few degrees. We assume here a range of approximately 10 degrees which corresponds to the value $\beta=0.5\,\text{K}^{-1}$.
As for the physical quantities associated with VO$_2$ they are~\cite{Mott,Barker} $\alpha_{1_d} = 26.4 \times 10^{-6}$\,K$^{-1}$, $\kappa_{1_d} = 3.6$\,W\,m$^{-1}$\,K$^{-1}$, $\epsilon_{1_d} = 0.8$ in the dielectric phase and $\alpha_{1_m} = 17.1 \times 10^{-6}$\,K$^{-1}$, $\kappa_{1_m} = 3.6$\,W\,m$^{-1}$\,K$^{-1}$, $\epsilon_{1_m} = 0.1$ in the metallic phase. Finally, the damping factor $\gamma$ is defined in terms of the oscillator quality factor $Q$ as~\cite{Springer} $\gamma = 3.52\, Q^{-1} L^{-2} \sqrt{EI\mu}$. It depends on the surrounding environment and on the first natural frequency of the cantilever. For a cantiver embedded in vacuum and in air the quality factors 
are $Q\approx100$ and $Q\approx50000$~\cite{Springer}, respectively.

\begin{figure}[t!]
\hspace{-0.12cm}\includegraphics[width=0.455\textwidth]{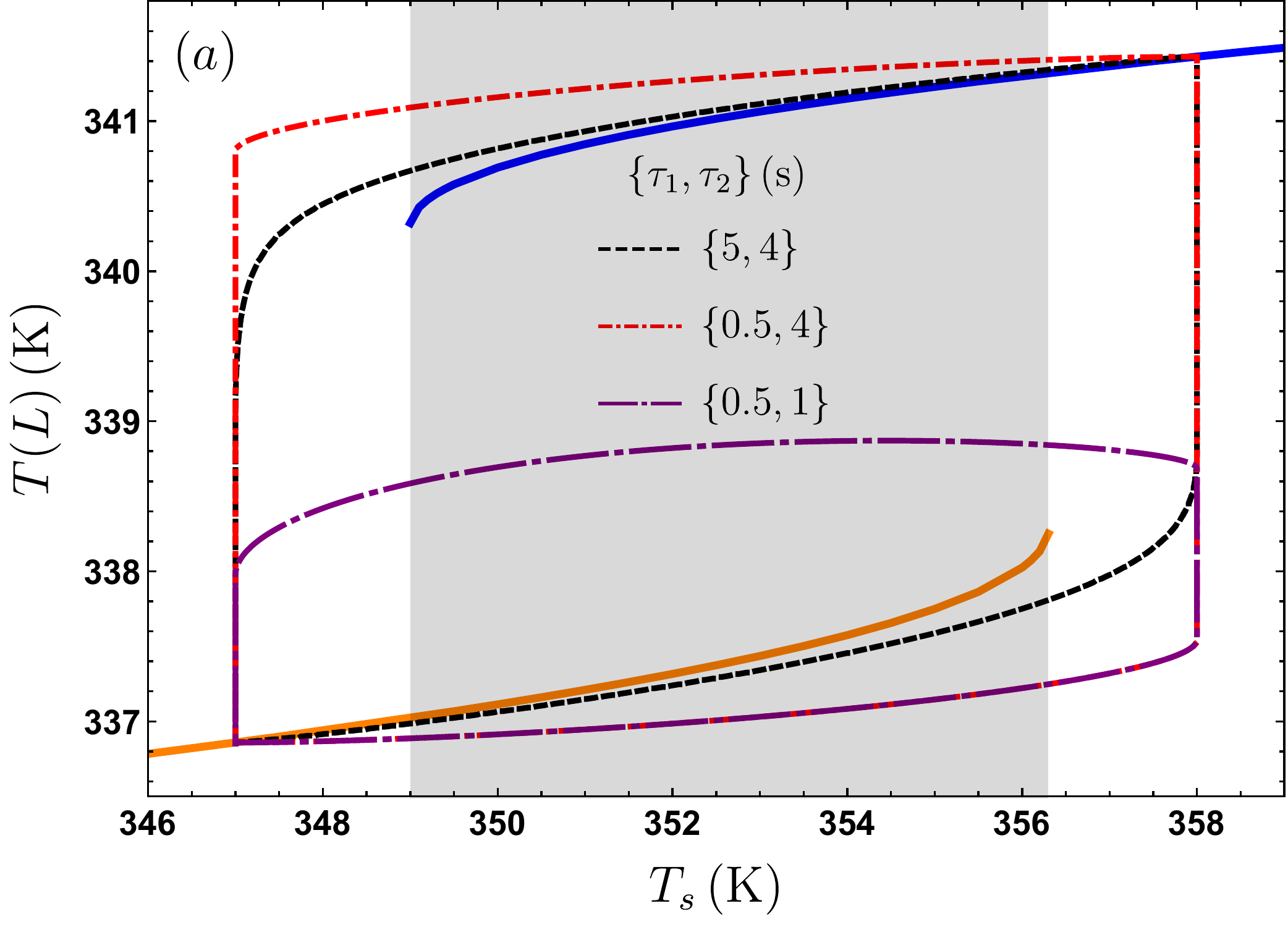}\\\vspace{0.3cm}
\hspace{-0.18cm}\includegraphics[width=0.455\textwidth]{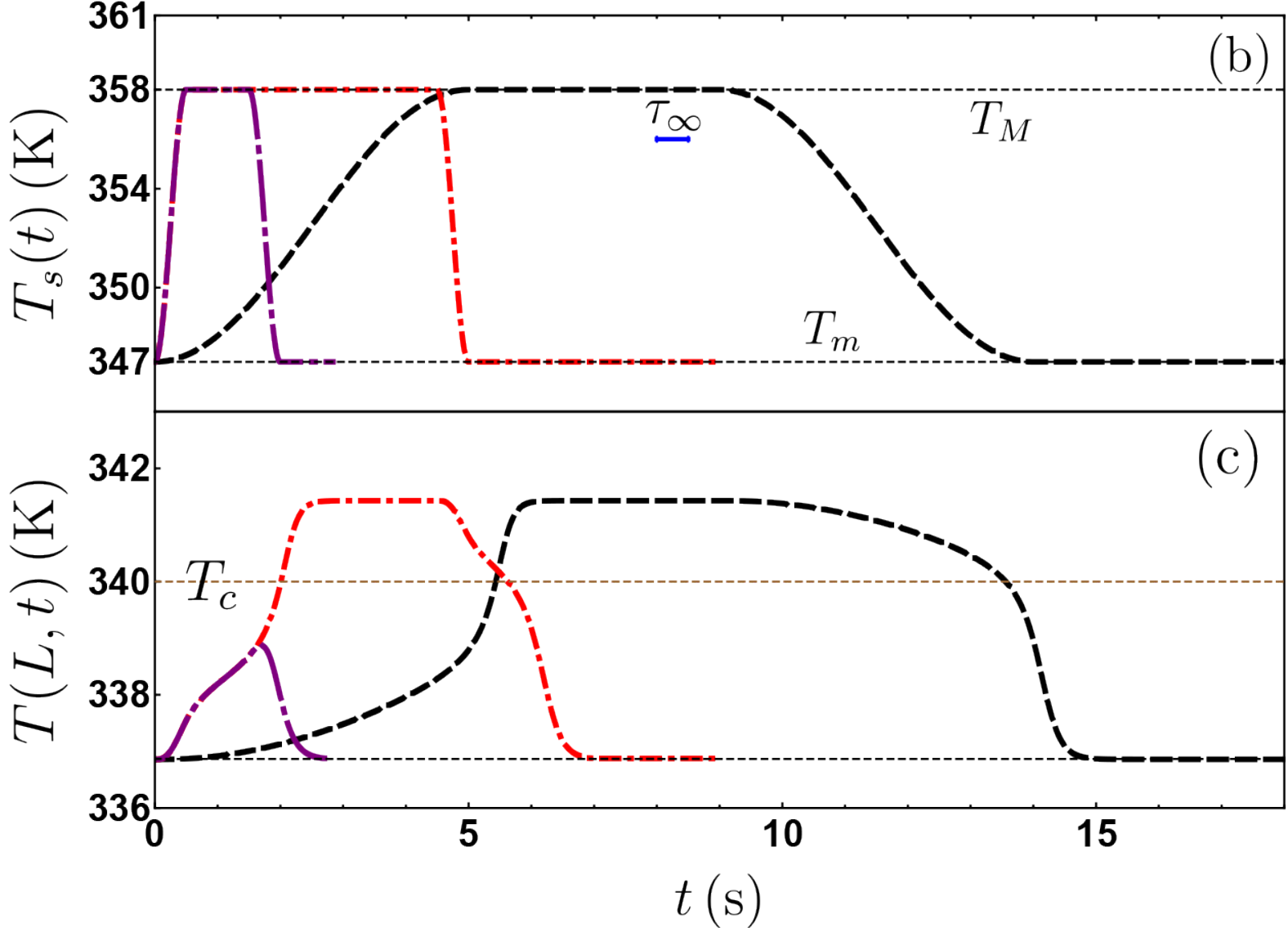}\\
\hspace{-0.0cm}\includegraphics[width=0.46\textwidth]{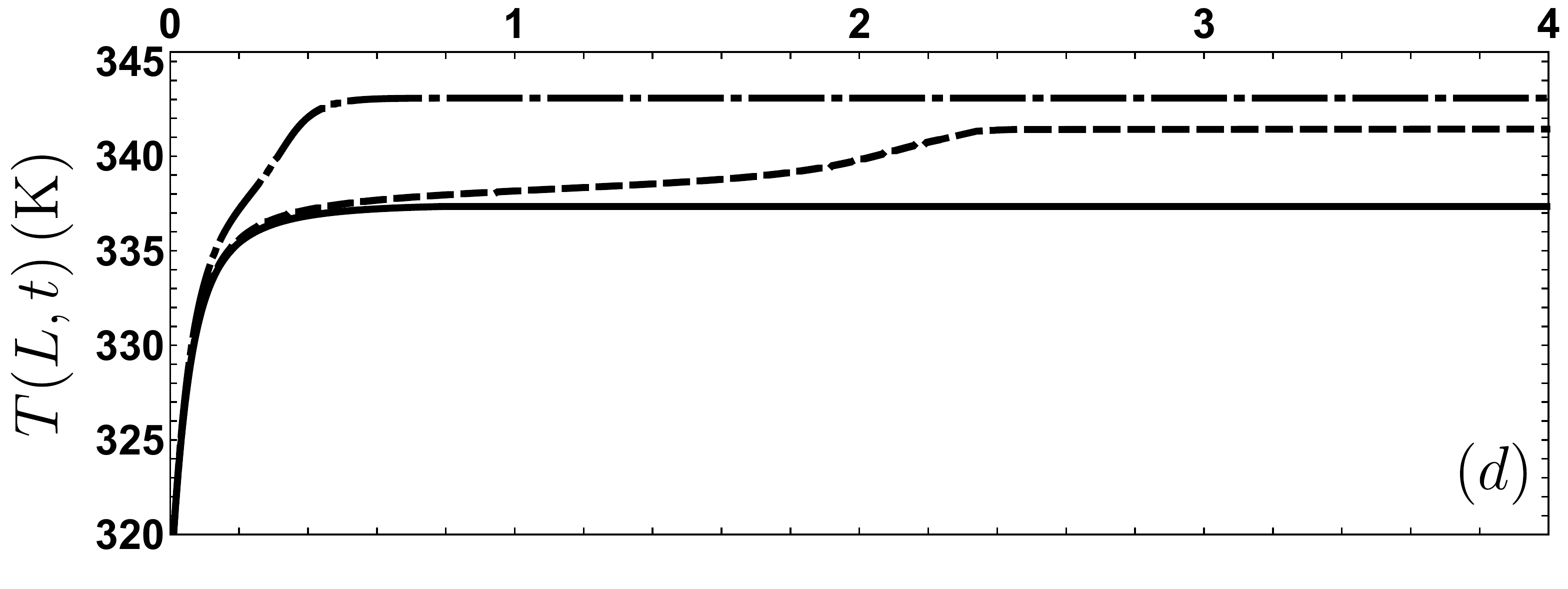}
\caption{(a) Phase diagram of the time-dependent temperature $T(L,t)$ of the free end of the cantilever as a function of the substrate temperature $T_s(t)$. The three different curves correspond to different profiles of $T_s(t)$ (see legend). The blue and orange solid curves correspond the steady-state solutions for $T(L)$ associated with each value of $T_s$. (b) Imposed time dependence of the substrate temperature $T_s$ (see text for functional dependence). The blue segment represents the decay timescale $\tau_c$. (c) Time evolution of $T(L)$ as a function of time [same color scheme as in panel (a)]. The horizontal brown dashed line represents the critical temperature $T_c$. (d) Time evolution of $T(L)$ for constant $T_s$ equal to 353\,K (solid), 358\,K (dashed) and 400\,K (dot-dashed).}
\label{fig:Hys}
\end{figure}

In Fig.~\ref{fig:t-Tu} we show the time evolution of $T(L,t)$ and $u(L,t)$ for the temperatures $\{T_e,T_w,T_s\}=\{300,356,353\}\,$K, with the boundary conditions $u(x,0)= g(x) = 135\,\text{m}^{-1} x^2$ and $T(x,0)= f(x)=\lambda x^2 - 2 \lambda L x + T_w$, where $\lambda$ is an adjustable parameter. Notice that a parabolic initial temperature profile is similar to the type of solution we get in steady-state regime. We observe in Fig.~\ref{fig:t-Tu} that for two different values of $\lambda$ [$\lambda = -0.1 \times 10^8$\,K\,m$^{-2}$ (blue curves) and $\lambda = 3 \times 10^8$\,K\,m$^{-2}$ (red curves)], that is for two different initial temperature profiles, the system evolves to two distinct stable solutions, thus demonstrating its bistability. More specifically, we show in the left inset of Fig.~\ref{fig:t-Tu}(a) that, as function of $\lambda$, only two stationary solutions for $T(L,t)$ are obtained. Since the surrounding medium only affects the oscillator damping and does not modify its asymptotic behavior, the steady-state solutions are the same both in vacuum and in air. We also observe that the steady state is reached after almost the same time interval ($\tau_\infty\approx 0.5\,$s) in both cases. This time is in agreement with the decay timescale $\tau_c = \rho\,C L^2 / \kappa \approx 0.14$\,s which can be extracted from a simple dimensional analysis of the energy-balance equation. The behavior of the displacement $u(L,t)$ is more interesting, and contrarily to the temperature evolution it clearly depends on the surrounding medium. Although the time needed both in air and in vacuum to reach the asymptotic regime is still $\tau_\infty$, the displacement oscillates differently in vacuum and in air. As shown in the left inset of Fig.~\ref{fig:t-Tu}(b), we clearly observe two oscillation frequencies, corresponding to the first and third natural frequencies of the cantilever, $\omega_1 = 3.52\,L^{-2} \sqrt{EI/\mu}$ and $\omega_3 = 61.7\,L^{-2} \sqrt{EI/\mu}$. Since the decay time $\tau_{\omega_i}$ of these oscillations is proportional to the quality factor ($\tau_{\omega_i} = 2Q/ \omega_i$) the longest decay timescales in air and vacuum are, respectively, $\tau^{\text{air}}_{\omega_1}=5.1\times10^{-3}$\,s, and $\tau^{\text{vacuum}}_{\omega_1}=2.5$\,s. As expected, the damping of the oscillations is stronger in air than in vacuum.

We now explore the possibility of using the bistability, identified so far only for a specific configuration, to produce a hysteretic behavior with respect to an external control parameter. A natural parameter is the substrate temperature $T_s$ or the wall temperature $T_w$ (modifiable tuned using for instance Peltier elements or external laser sources). Here we will focus on $T_s$ as a control parameter. We start by identifying numerically the range for this temperature over which the bistability is present. To this aim, we perform numerical calculations keeping all the parameters, except $T_s$, unchanged. The steady solutions for $T(L)$ are represented by the blue and orange solid lines in Fig.~\ref{fig:Hys}(a). We clearly see that, when $T_s$ lies in the range $[349,356.3]\,$K [gray zone in Fig.~\ref{fig:Hys}(a)], we observe a bistable behavior, while for values of $T_s$ outside this range we only get one stable solution. As suggested by the curves plotted in Fig.~\ref{fig:Hys}(a), a time variation of $T_s$ allows to switch from one stable solution to the other one through a hysteresis loop. Of course, this possibility strongly depends on the specific time dependence of $T_s$, and in particular on the comparison between the typical timescale over which $T_s$ is tuned and the relaxation time $\tau_c$ of our system. To get a deeper insight into this aspect, we let $T_s$ vary according to the time-dependent function represented in Fig.~\ref{fig:Hys}(b). We start from a minimum temperature $T_m = 347$\,K and from $t=0\,$s to $t=\tau_1$ we increase the value of $T_s$ up to its maximum $T_M=358$\,K through the growing branch of the function $(T_M - T_m)(1-\cos[\pi\,t/\tau_1])/ 2$. We then keep $T_s = T_M$ during a time interval $\tau_2$ to finally go down to $T_m$ over a time interval $\tau_1$ through the descending branch of the function $(T_M - T_m)(1-\cos[\pi(t - \tau_2 )/\tau_1])/2$.

In Fig.~\ref{fig:Hys} we describe the evolution of $T(L)$ as a function of $T_s$ when this one is modulated at different timescales [see Fig.~\ref{fig:Hys}(b)]. When the period of modulation is smaller than the time of thermal relaxation of the cantilever, the latter is not sufficiently heated up to transit into its metallic phase [dot-long-dashed curve in Fig.~\ref{fig:Hys}(c)]. On the contrary, with a slower thermal excitation the system is able to perform the transition and its temperature switches from the lower stable solution to the upper one beyond $T_c$. 

The two stable solutions plotted in Fig.~\ref{fig:Hys}(a) correspond to a net heat flux which is locally convex in the ($T_s,T(L)$) plane. In Fig.~\ref{fig:Hys}(d) we also demonstrate the presence of an unstable solution. The time evolution of $T(L,t)$ is plotted for three fixed values of $T_s=353,358,400$\,K, assuming that the initial temperature profile is $T(x,0)=f(x)$ with $\lambda = 3 \times 10^8$\,K\,m$^{-2}$. For $T_s=353\,$K ($T_s=400\,$K) we observe the expected convergence to the lower (upper) solution on a timescale $\approx\tau_c$. Differently, for $T_s=358\,$K we highlight an intermediate plateau before the system converges to its unique steady-state solution. This is a signature of the fact that $T_s$ is still close to the region where two solutions exist, and is analogous to the saddle-point behavior already observed in Ref.~\cite{pba2}. 

A direct application of this thermomechanical bistability is the thermal treatment of information. It is straightforward to see that such a system can operate as a NOT gate when the control parameter $T_s$ is the input of the gate and $T(L)$ its boolean output. If we define a thermal state `0' as the state where $T_s$ is close to $T_{m}$ and a thermal state `1' as the state where $T_s$ is close to $T_{M}$ [by defining appropriate treshold temperatures] and on the other hand two states `1' and `0' when $T(L)<T_{c}$ (larger bending) and $T(L)>T_{c}$ (smaller bending), then the cantilever behaves like a NOT gate. The coupling of such oscillators and their control with multiple input parameters could allow to define more complex logical operations.

We have shown that a phase-transition cantilever in a scenario out of thermal equilibrium may have a bistable thermomechanical behavior. We have demonstrated that its temperature profile can be driven by external heat flux and switched from one stable state to another paving thus the way to basic logical operations using external thermal control parameters. Several open questions remain to be explored at a fundamental level, including the thermal prepation of oscillators, their coupling with other oscillators and their scalability to operate at different time and spatial scales.

\begin{acknowledgments}
The authors are grateful to Prof. Pietro Salvini for very fruitful discussions. 
\end{acknowledgments}

\end{document}